\documentclass[runningheads]{llncs}

\usepackage[utf8]{inputenc} 
\usepackage[T1]{fontenc}    
\usepackage{hyperref}       
\usepackage{url}            
\usepackage{booktabs}       
\usepackage{amsfonts}       
\usepackage{nicefrac}       
\usepackage{microtype}      
\usepackage{lipsum}
\usepackage{threeparttable}
\usepackage{multirow}

\usepackage[pdftex]{graphicx} 

\title{Radiologist-level stroke classification on non-contrast CT scans with Deep U-Net}

\titlerunning{Radiologist-level stroke classification on non-contrast CT scans}
\author{Manvel Avetisian\inst{1} \and
Vladimir Kokh\inst{1} \and
Alex Tuzhilin\inst{2} \and
Dmitry Umerenkov\inst{1}}
\authorrunning{Avetisyan M., Kokh V., Tuzhilin A. and Umerenkov D. }
\institute{Sberbank AI Lab
\email{avetisyan.m.s@sberbank.ru}, \email{kokh.v.n@sberbank.ru}, \email{d.umerenkov@gmail.com} \and
New York University
\email{atuzhili@stern.nyu.edu}}

\begin{document}
\maketitle

\begin{abstract}
Segmentation of ischemic stroke and intracranial hemorrhage on computed tomography is essential for investigation and treatment of stroke. In this paper, we modified the U-Net CNN architecture for the stroke identification problem using non-contrast CT. We applied the proposed DL model to historical patient data and also conducted clinical experiments involving ten experienced radiologists. Our model achieved strong results on historical data, and significantly outperformed seven radiologist out of ten, while being on par with the remaining three.
\end{abstract}


\section{Introduction}
A stroke is a serious medical problem consisting of two types: ischemic stroke (IS) that occurs if an artery supplying blood is blocked, and intracranial hemorrhage (IH) that occurs when an artery leaks blood or ruptures \cite{saver2013time}. Among various stroke detection methods, non-contrast computed tomography (CT) is of particular importance because of it's speed of imaging, widespread availability and possibility of using in an emergency cases. One of the main problems with non-contrast CT scans is that it is hard to accurately detect cases of acute stroke \cite{schriger1998cranial}, \cite{chalela2007magnetic}. Moreover, there is a shortage of skilled professionals capable of reliably reading CT scans, especially in rural and remote areas. Furthermore, even experienced radiologists make mistakes, each of that mistakes possibly resulting in serious trauma or even death.
In this paper we propose a novel deep-learning based method that reliably identifies the existence of a stroke, the type of the stroke and the localization area of the stroke. We tested our method on historical data and also in a controlled experimental setting. We demonstrate that our method produced strong performance, as will be shown in the paper. Presented segmentation-based classifier can be used as a radiologist tool by classifying patients with supporting evidence provided by the segmented probable stroke area. 

The contributions of this paper are as follows. First, we applied deep neural networks to the task of segmentation of IS areas using non-contrast CT images. Second, we applied our method to 300 CT scans annotated by experienced radiologists, and demonstrated strong performance on this data (0.703 Dice Similarity Coefficient (DSC) in three-way segmentation of healthy tissue, IS and IH). Third, we created a three way classifier based on our segmentation model and did a clinical experiment with experienced radiologists involving 180 cases. Our model significantly outperformed seven out ten radiologists while being on par with the remaining three.

\section{Related work}
Neural networks (NN) have outperformed doctors on several medical imaging tasks, including those in the fields of dermatology \cite{esteva2017dermatologist} and ophthalmology \cite{gulshan2016development}. While CT scan data is widely used for the task of automated segmentation and classification of other organs, the overwhelming majority of NN approaches to brain image analysis are done using MRI data \cite{litjens2017survey}.

Recently there have been several projects launched on applying NN to non-contrast head CT scans. In \cite{chilamkurthy2018deep} authors used NN algorithms for detection of critical findings including IH in non-contrast CT scans. In \cite{ramos2018convolutional} researchers applied Convolutional Neural Networks (CNN) to the task of segmentation of IH and edema,  achieving DSC of 0.74 for IH segmentation. However, \cite{ramos2018convolutional} used only images of patients with confirmed IH, which makes their model not useful as a detection tool in clinical setting.  
In \cite{lisowska2017context}, authors applied a CNN to the task of IS detection, achieving ROC-AUC of 0.915 on the hemisphere level. However, while authors did voxel-level classification, they focused not on refining the detection boundary but on the final classification accuracy, rendering their method unsuitable for the segmentation task. Moreover, \cite{pereira2018stroke} applied shallow CNN for the three-way stroke classification. However, the size of the dataset used in \cite{pereira2018stroke} was very small, limiting the statistical significance of the achieved results. In particular, they used 100 slices for each type, which is only 1\% of the dataset size used in this study. 
Also, \cite{abulnaga2018ischemic} \cite{dolz2018dense} used U-net based CNNs for segmentation of IS lesions using computed tomography with perfusion (contrast). Contrary to \cite{abulnaga2018ischemic} \cite{dolz2018dense}, we did it using non-contrast CT scans. 

Despite all this work, nobody applied NNs to the task of segmentation of IS areas using non-contrast CT images, tested them on large datasets and achieved stroke detection results surpassing experienced radiologists.
\section{Data}
\subsection{Dataset}
\label{dataset}
Our labeled dataset includes retrospective anonymized non-contrast head CT scans in the following conditions: acute IS, IH and the healthy tissue. These images have been taken on different CT scanners of three different hospitals. All the personally identifiable information has been deleted by the hospitals. The total number of scans is 300 (200 with IS, 100 with IH); they were obtained in the acute period and confirmed clinically. Timing-wise the range of cases varies significantly from the early onset of the stroke (a matter couple of hours) to the extensive stroke (over 10 hours,  including wake-up strokes).
Each case was analyzed by three independent radiologists with at least ten years of extensive experience in diagnosing acute cerebral strokes. Furthermore, they had access to the patient clinical records. In each of the 300 studies, the radiologists manually labeled the stroke areas by isolating the area of interest in each slice of 5 mm for IS, and 3 mm for IH. We split our dataset into the training and validation sets in the proportion of 90\% and 10\%. 
\subsection{Data Preparation, Interpolation and Augmentations}
For each patient, the data was received from radiologists in the DICOM format and contained at least three data series: low slice thickness (0.5mm - 1mm), unlabeled; high slice thickness (3-5mm), unlabeled; high slice thickness (3-5mm) with labeled contours of stroke areas.
The contours where labeled on the high but not on the low thickness slices to reduce manual work of radiologists. Unfortunately, this approach greatly reduces the training data. Therefore, we applied an linear interpolation algorithm to create a set of labeled low thickness slices.
The original high slice thickness images contained 11,919 slices, including 3,173 slices with stroke areas (1,623 hemorrhagic and 1,550 ischemic). Interpolated low thickness images contained 81,642 slices, including 19,726 slices with stroke areas (12,596 hemorrhagic and 7,130 ischemic).
We used the DICOM's rescale intercept and rescale slope tags to normalize raw voxel values. 

We used data augmentation technique \cite{krizhevsky2012imagenet} to increase training data variability and improve network generalization because image transformations have been empirically proven effective in the image recognition tasks \cite{krizhevsky2012imagenet}, \cite{he2015delving}. In this work we used random 90 degree rotations, random flips with probability 50\%, random size shift in range 90-110\% and random rotations up to 45 degrees.

\section{Method}
\subsection{The U-Net based Model}
The vast majority of work in medical image segmentation constitutes variations of the U-net architecture \cite{ronneberger2015u} that has several well-known advantages \cite{litjens2017survey}. As a modification of the original U-net, we used a state-of-the art image classification architecture, as the encoder part instead of the VGG-like encoder used in the original paper \cite{ronneberger2015u}. Another modification to the architecture described in \cite{ronneberger2015u} is the use of asymmetric decoder containing less parameters than the encoder. We did it to limit model complexity. 

We used a recent classification architecture Dual Path Network (DPN) \cite{chen2017dual}, which combines the features of residual networks \cite{he2016deep} and densely-connected networks \cite{huang2017densely} with depth 94 as the encoder architecture. We did several experiments using an encoder pretrained on the ImageNet dataset \cite{imagenet_cvpr09} with and without freezing the encoder weights; however it did not help us to produce better results, and therefore we decided to train our model from scratch. We used the same image as the input for the three channels of the pretrained classification encoder.

We build our final model (DPN92-Unet) using a slim decoder, as presented in Fig. 1. Each decoder stage consists of a 3x3 convolution layer with 16 input and output channels followed by batch normalization layer and ReLU nonlinearity. Convolutional bottlenecks with 3x3 kernel and 16 output channels were used to fuse encoder path with decoder path. We have also upsampled the output of all decoder blocks as needed and stacked into a single hypercolumn \cite{hariharan2015hypercolumns}. Furthermore, we used this hypercolumn to make final pixel-wise predictions by a depthwise convolution with 2 output channels, corresponding to probabilities of this pixel belonging to ischemic area or hemorragic area. We also experimented with adding squeeze and excitation blocks \cite{hu2018squeeze} after each encoder stage to improve contextual processing. However, our experiments showed that squeeze and excitation block didn't work in our case.
We used a batch norm \cite{ioffe2015batchnorm} as a means to regularize our neural network.
\begin{figure} [ht]
 \caption{Model architecture}
 \includegraphics[width=\textwidth]{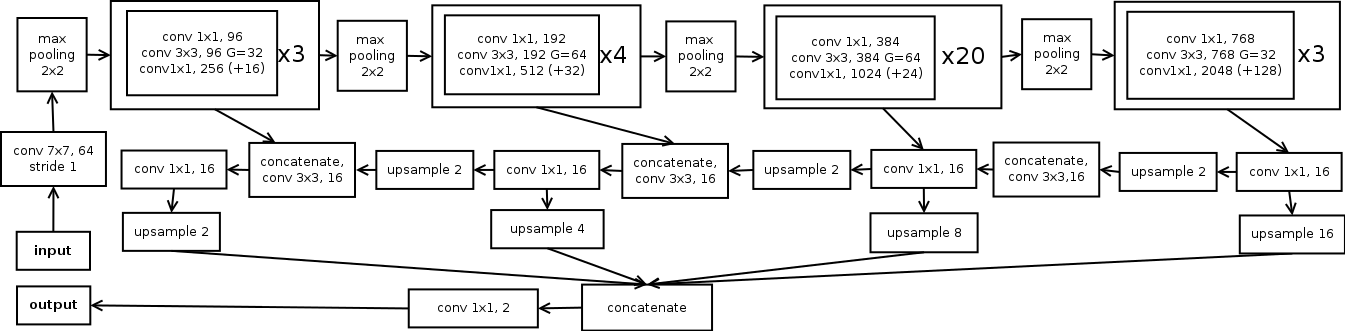}
 \label{fig-reg}
\end{figure}
\subsection {Training}
Model training was done for 20,000 steps on the training set described in Section \ref{dataset}. At each step, a batch of 12 slices was processed, slices for each batch being selected with probability of 50\% randomly from slices with stroke regions and with probability 50\% from all the available slices. In one experiment we also tried to add additional CT images of healthy tissue from a separate dataset of 100 patients to increase the negative class variability, but without much success.
We used Focal Loss \cite{lin2017focal} as the loss function and RMSProp \cite{hinton2012neural} as the optimizer with starting learning rate of 0.0001 and reduced it by the factor of 0.99977 at each step.
We selected the best weights based on the average Intersection over Union (IoU) metric calculated for each patient in the validation set. This statistics was calculated every 500 steps.
\subsection{Test time augmentations and ensembling}
As a further way to increase model accuracy (at the expense of prediction time) we experimented with applying test time augmentations (TTA) by averaging the predictions of the model on the original slices and the slices after applying left-right flips and up-down flips. However, we did not manage to get statistically significant segmentation quality improvements using this method.
The strategy that worked for us to increase the segmentation score was to (a) train multiple models described above (about 40), (b) select several best models according to IoU metric (in our case 6 out of 40), and (c) average their prediction results.
\subsection{Prediction in 3 projections and segmentation postprocessing}
To take advantage of the three-dimensional nature of the data, we created two additional datasets with the coronal and sagittal projections to complement the original dataset in axial projections. For coronal and sagittal projections we averaged the results of two independently trained models. This was done due to time constraints in training the model and in a an effort to avoid increasing the prediction time too much. While the models trained on these additional datasets achieved lower segmentation quality by themselves, combining predictions in all the three projections with different weights led to a noticeable increase in segmentation quality, as will be demonstrated in Section 5. 

After obtaining the predictions for all the three projections, we consider voxels, where in at least one of the projections the model output is greater that corresponding coefficient (k\textsubscript{axial}, k\textsubscript{coronal}, k\textsubscript{sagittal}) to be stroke voxels. We selected the coefficients to maximize the average DSC on the validation set for the three class segmentation.
\[ k_{axial}=0.47, k_{coronal}=0.56, k_{saggital}=0.56   \]
After combining the three projections into a binary prediction, we applied morphological closing \cite{serra1983image} with a radius 3 ball-shaped structuring element to produce the final result.
\subsection{Segmentation based classifier}
We used the segmentation results described in the previous section to classify the whole head CT images into the following three classes: healthy, acute IH and acute IS. As the classification criteria, we calculated the following quantity for each voxel:
\[ V_{pred} = Pred_{axial}/k_{axial} + Pred_{coronal}/k_{coronal}  
+ Pred_{saggital}/k_{saggital}   \]
The coefficients used are the same as for the voxelwise segmentation described in the previous section.
After that, for each patient, we calculated the mean voxel value for all the voxels with value greater than 0.9. Images with mean ischemic value greater than 1.16 where assigned to the ischemic cases, and images with mean hemorrhagic value greater than 1.52 where assigned to the hemorrhagic cases. If neither of those criteria were true, we assigned the image to the no-pathology case. The classification constants were tuned on a separate dataset containing classification results for 30 cases for each of IS, IH and healthy tissue. 
\section{Results}
\subsection{Segmentation}
We evaluated performance of our segmentation method presented in Section 4 on historical data using DSC measure, and the results are presented in Table 1. Each row in Table 1 corresponds to the approaches previously presented in Section 4.1 on the projections described in Section 4.4. The first column in Table 1 shows the results for the subset of IH from the validation dataset, the second column corresponds the IS subset of validation dataset, and the third column IS+IH corresponds to the full validation dataset. As Table 1 shows, we indeed improved the performance results by using ensembling, combining predictions from the three projections and postprocessing. In particular, the last line of Table 1 shows the performance result of 0.703 for the three-projections model,  which is significantly better than the performance results of the single projection models (with $p < 0.05$ in all the cases).
\begin{table}[ht]
\begin{threeparttable}
\caption{Performance results of segmentation on validation datataset for different stroke types measured by mean patientwise DSC and std}
\begin{tabular}{ | l | c | c | c | c }
\hline
\textbf & \textbf{IH} & \textbf{IS}& \textbf{IH+IS} \\
\hline
\textbf{Axial projection} & & & \\
{DPN92-Unet}\tnote{1} & 0.669 (0.003)& 0.609 (0.012) & 0.629 (0.003) \\
{DPN92-Unet, encoder pretrained}\tnote{1} & 0.652 (0.021) & 0.611 (0.006) & 0.625 (0.005) \\
{DPN92-Unet, encoder pretrained + warmup}\tnote{1} & 0.625 (0.035) & 0.542 (0.045) & 0.570 (0.037)  \\
{DPN92-Unet, additional healthy examples}\tnote{1} & 0.615 (0.026) & 0.600 (0.023)& 0.605 (0.012)  \\
{DPN92-Unet, encoder pretrained, SSCE}\tnote{1} & 0.654 (0.025) & 0.618 (0.017)& 0.630 (0.009)  \\
{DPN92-Unet with TTA}\tnote{1} & 0.683 (0.021) & 0.606 (0.015)& 0.632 (0.008) \\
{Ensemble of 6 best DPN92-Unets} & 0.675  & 0.634 & 0.648   \\
{OOF Ensemble on new dataset}\tnote{3} & 0.557\tnote{3} & 0.524\tnote{3} & 0.616\tnote{3}   \\
\hline
\textbf{Coronal projection} & & &  \\
{DPN92-Unet}\tnote{2} & 0.547 & 0.405 & 0.457  \\
\hline
\textbf{Sagittal projection} & & &  \\
{DPN92-Unet}\tnote{2} & 0.623 & 0.425 & 0.491  \\
\hline
\textbf{Three projections} & & &  \\
{Three DPN92-Unet ensembles} & 0.730 & 0.668 & 0.689\\
{Three DPN92-Unet ensembles + postprocess} & 0.732 & 0.689 & 0.703\\
\hline
\end{tabular}
\begin{tablenotes}\footnotesize
\item[1] Results are averaged for the best 5 out of 6 independently trained identical models with different random initializations
\item[2] Results are averaged for two identical models independently trained with different random initializations.
\item[3] DSC are not comparable with other results in this table. Results are for out-of-fold training for 6 folds on original dataset with additional training cases, selected by radiologists based on model errors made during the clinical experiment.
\end{tablenotes}
\label{tab-loss}
\end{threeparttable}
\end{table}
As we can see from Table 1, the cases of IH are segmented better than the cases of IS, even though the amount of training cases for IS was twice as high as for IH. Therefore, we conclude that the task of segmentation of IS on non-contrast CT is significantly harder than the task of segmentation of IH, which is consistent with the performance of human radiologists, as described in Section 5.2.

\subsection{Clinical experiment}
We conducted a clinical experiment based on the retrospective cases, in collaboration with one of the major medical research center located in a European country. One hundred eighty non-contrast head CT scans were selected in various hospitals for our experiment (hereinafter ExpData). The collected ExpData included 60 cases for each diagnosis - IS, IH and healthy tissue.
The process of data selection and confirmation of diagnosis was made by a group of three experienced radiologists, having at least ten years of continuous and extensive experience in the acute stroke brain diagnosis. As a result, each case had the diagnosis collectively made by a group of three of these experts. All the collected cases were blinded by removing any patient specific information and shuffled randomly.
In this experiment, ten radiologists from five leading hospitals in the region were independently tasked to analyze these cases and to make diagnosis. The radiologists had no access to the patient clinical records of the analyzed cases and did not know about the ratio of IS / HS / healthy tissue in ExpData. In our experiment, these radiologists used the DICOM imaging analysis software, and had no time limit for the analysis.
The experimental process was controlled by an independent researcher to achieve objective results.

\renewcommand{\arraystretch}{1.3}
\begin{table}[ht]
\caption{Comparative analysis of errors of our model vs radiologists (out 180 cases)}
\begin{tabular}{ | c || c | c | c | c | c | c | c | c | c | c | c |}
\hline
 & \multicolumn{10}{|c|}{Radiologist} & \multirow{2}{*}{Our model}\\
\cline{2-11}
 & 1 & 2 & 3 & 4 & 5 & 6 & 7 & 8 & 9 & 10 &  \\
\hline
\hline
Errors & 6 & 15 & 8 & 19 & 33 & 23 & 25 & 18 & 19 & 19 & 7 \\
\hline
p-value & 0.999 & 0.122 & 0.999 & 0.024 & 0.001 & 0.004 & 0.001 & 0.037 & 0.024 & 0.024 &  \\
\hline
\end{tabular}
\label{tab-loss1}
\end{table}

We compare the performance of each radiologist against performance of our model using standard accuracy metric, and the Fisher's exact test was used to calculate the p-value. We conclude from this analysis that our model significantly outperforms 7 radiologists out of 10 and achieves on par results with the remaining 3 radiologists (see Table 2 for specifics). Segmentation quality is not compared as labels are not available for the ExpData.

\section{Conclusions}
In this paper we proposed a novel approach to segmentation and classification based on the U-Net architecture that we modified in various ways to better perform acute stroke segmentations. We conducted an experiment on historical data and also did a clinical experiment involving ten experienced radiologists who analyzed 180 CT scan cases. On the historical data, our model made only 7 mistakes out of 180 and significantly outperformed 7 radiologists out of 10, while being on par with the remaining 3 radiologists. As a future work, we are planning to improve our system even further so it would outperform radiologists even better and make even fewer mistakes in the most difficult cases, including the cases of the early onset of strokes (a few hours). We are also planning to extend our approach to other diseases (e.g. traumatic brain injury, brain tumors).

\bibliographystyle{unsrt}  

\end{document}